\documentclass[conference]{IEEEtran}
\IEEEoverridecommandlockouts
\usepackage[binary-units]{siunitx}
\usepackage{tabularx}
\usepackage{multirow}
\usepackage{color, colortbl}
\usepackage[binary-units]{siunitx}
\DeclareSIUnit{\bps}{bps}
\usepackage{graphicx}
\usepackage{cite}
\usepackage{amssymb}
\usepackage{amsmath}
\usepackage[caption=false,font=footnotesize]{subfig}

\begin{document}
\title{Cell-Edge Performance Booster in 6G: \\ Cell-Free Massive MIMO vs. Reconfigurable Intelligent Surface }

\author{\IEEEauthorblockN{Wei Jiang\IEEEauthorrefmark{1} and Hans D. Schotten\IEEEauthorrefmark{2}}
\IEEEauthorblockA{\IEEEauthorrefmark{1}German Research Center for Artificial Intelligence (DFKI)\\Trippstadter Street 122,  Kaiserslautern, 67663 Germany\\
  }
\IEEEauthorblockA{\IEEEauthorrefmark{2}Rheinland-Pf\"alzische Technische Universit\"at Kaiserslautern-Landau\\Building 11, Paul-Ehrlich Street, Kaiserslautern, 67663 Germany\\
 }
%\thanks{This work was supported in part by the European Commission H2020 Framework Programme through \textit{AI@EDGE} (Grant no. \emph{101015922}) and in part by the German Federal Ministry of Education and Research (BMBF) through \emph{Open6G-Hub} (Grant no.  \emph{16KISK003K}).}
}

\maketitle

\begin{abstract}
%\boldmath
User experience in mobile communications is vulnerable to worse quality at the cell edge, which cannot be compensated by enjoying excellent service at the cell center, according to the principle of risk aversion in behavioral economics. Constrained by weak signal strength and substantial inter-cell interference, the cell edge is always a major bottleneck of any mobile network. Due to their possibility for empowering the next-generation mobile system, reconfigurable intelligent surface (RIS) and cell-free massive MIMO (CFmMIMO) have recently attracted a lot of focus from academia and industry. In addition to a variety of technological advantages, both  are highly potential to boost cell-edge performance.  To the authors' best knowledge, a performance comparison of RIS and CFmMIMO, especially on the cell edge, is still missing in the literature. To fill this gap, this paper establishes a fair scenario and demonstrates extensive numerical results to clarify their behaviors at the cell edge.
\end{abstract}

\IEEEpeerreviewmaketitle

\section{Introduction}
In a cellular network, a base station (BS) is placed at the center of a cell of a network of cells and   serves a lot of users simultaneously.
%With simple signal processing, it can provide high throughput, reliability, and energy efficiency. The massive number of service antennas in a cell can be deployed in collocated or distributed setups. Collocated massive MIMO architectures, where all service antennas are located in a compact area, have low backhaul requirements and joint processing. Nonetheless,
High quality of service (QoS) is primarily offered to users close to the BS, namely the cell center. The area of a cell center is only a small portion of the whole coverage, according to a simple geometry calculation. Many users are at the cell edge, suffering from worse QoS due to weak signal strength, which is restricted by large distance-dependent path loss, strong inter-cell interference, and handover issues that are inherent to the cellular architecture.  The performance gap between the  center and edge in a cell is not trivial; however, it is tremendous.  ITU-R  M.2410  \cite{Ref_non2017minimum} specifies the minimum requirements related to  technical performance for IMT-2020 (the official naming of 5G), where peak spectral efficiency (SE) in the downlink and uplink  reaches \SI{30}{\bps\per\hertz^{}} and \SI{15}{\bps\per\hertz^{}}, respectively. The so-called $5^{th}$ percentile SE,  a.k.a $95\%$-likely SE, can be guaranteed to $95\%$ of the users and thus define user-experienced QoS. In contrast, the targets of $5^{th}$ percentile SE are only \SI{0.3}{\bps\per\hertz^{}} (downlink) and \SI{0.21}{\bps\per\hertz^{}} (uplink) in indoor hotspot enhanced mobile broadband (eMBB), which are further lowered to \SI{0.12}{\bps\per\hertz^{}} and \SI{0.0453}{\bps\per\hertz^{}} in rural eMBB, amounting to  a huge difference more than 100 times.

Recently, two novel techniques - cell-free massive MIMO (CFmMIMO) and reconfigurable intelligent surface (RIS) - have attracted a lot of focus from academia and industry due to their potential for the Sixth Generation (6G) system \cite{Ref_jiang2021road}.
CFmMIMO \cite{Ref_ngo2017cellfree} is a distributed massive MIMO system, where a large number of antennas serve a few users spread over a wide area. All distributed antennas cooperate phase-coherently via a fronthaul network and serve all users in the same time-frequency resource. There are no cells or cell boundaries, completely addressing the issues of cell edge and handover. RIS is a planar surface consisting of a large number of passive elements,  each of which is capable of independently inducing a phase shift to an impinging signal \cite{Ref_renzo2020smart}. In contrast to traditional wireless techniques, which can only \textit{passively} adapt to a wireless channel, RIS \textit{proactively} reshapes it to realize an on-demand propagation environment \cite{Ref_wu2019intelligent}. By collaboratively adjusting the phase shift of each element, the reflected signals combine constructively with the direct signal for signal amplification, or destructively for interference suppression.

In addition to a variety of technological advantages, both RIS and CFmMIMO are high potential to boost cell-edge performance.  To the authors' best knowledge, a performance comparison of RIS and CFmMIMO, especially on the cell edge, is still missing in the literature. To fill this gap, this paper establishes a fair comparison scenario and provides extensive numerical results to clarify their respective pros and cons. The remainder of this paper is organized as follows: Sections II and III provide a brief and informative introduction to CFmMIMO and RIS, respectively. In Section IV, the simulation setup and numerical results are explained to compare them. Section V concludes this paper.

\section{Cell-Free Massive MIMO}
\begin{figure}[!t]
    \centering
    \includegraphics[width=0.42\textwidth]{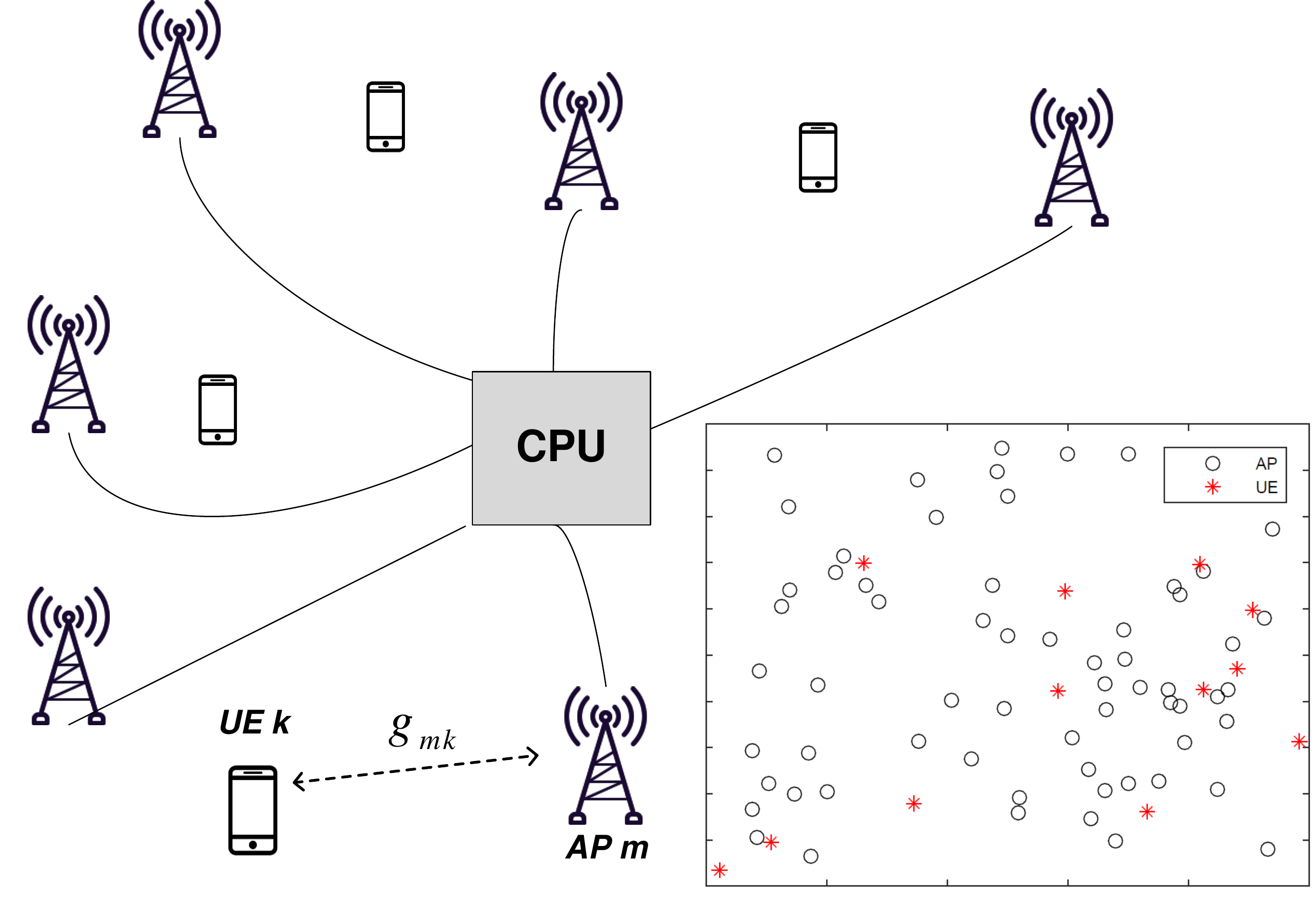}
    \caption{Schematic diagram of a cell-free massive MIMO system, where a large number of distributed APs serve a few users under the control of a CPU. The right-bottom figure shows an example layout of this system. }
    \label{fig:CFsystem}
\end{figure}

In a CFmMIMO system, a large number of $M$ single-antenna access points (APs) randomly spreads over a geographical area. These APs are coordinated by a central processing unit (CPU) via a fronthaul network to simultaneously serve $K$ user equipment (UE) over the same time-frequency resource, as shown in \figurename \ref{fig:CFsystem}. Time-division duplex (TDD) is applied to alleviate the high overhead of downlink pilots, which is proportional to the number of APs \cite{Ref_marzetta2015massive}. Write $g_{mk}=\sqrt{\beta_{mk}} h_{mk}$ to denote the channel gain between AP $m$, $\forall m$ and UE $k$, $\forall k$, where $\beta_{mk}$ and $h_{mk}$ represent large-scale and small-scale fading, respectively.
In the downlink of CFmMIMO, conjugate beamforming (CBF) and zero-forcing precoding (ZFP) \cite{Ref_jiang2021impactcellfree} are two simple but effective schemes that can spatially multiplex the information symbols $\textbf{u}=[u_1,\ldots,u_K]^T$, where $\mathbb{E}[|u_k|^2]=1$,  intended for $K$ users.

The communication process of CBF-based CFmMIMO is:
\begin{enumerate}
\item Each AP measures $\beta_{mk}$ and reports to the CPU for computing power-control coefficients $\eta_{mk}$. Usually, $\beta_{mk}$ keeps constant for a relatively long period.
\item Users synchronously send their pilot signals so that each AP estimates its channel signature $  \hat{\textbf{g}}_m=\bigl[\hat{g}_{m1},\hat{g}_{m2},\ldots,\hat{g}_{mK} \bigr]^T$.
\item AP $m$ locally precodes the information symbols as \begin{equation}
    s_m= \sqrt{ P_m}\sum_{k=1}^K\sqrt{\eta_{mk}}\hat{g}_{mk}^*u_k,
\end{equation}
where $P_m$ is the power limit of AP $m$.
\item APs synchronously send their transmitted symbols.
\end{enumerate}

Thus, the $k^{th}$ user observes
\begin{align} \nonumber
    y_k &=  \sum_{m=1}^M g_{mk}s_m +n_k = \underbrace{ \sum_{m=1}^M \sqrt{P_m\eta_{mk}} g_{mk} \hat{g}_{mk}^*u_k}_{\text{Desired\:Signal}} \\ \nonumber
      &+ \underbrace{ \sum_{m=1}^M   \sqrt{ P_m}\sum_{i\neq k}^K\sqrt{\eta_{mi}}g_{mk}\hat{g}_{mi}^*u_i}_{\text{Multi-User\:Interference}}+n_k,
\end{align}
where $n_k$ is additive white Gaussian noise (AWGN) with zero mean and variance $\sigma_n^2$, i.e., $n_k\sim \mathcal{CN}(0,\sigma_n^2)$.
Without downlink pilots, each UE detects its received signal based on the channel \textit{statistics}, i.e.,  $\mathbb{E}\left[\left \vert \hat{g}_{mk} \right \vert ^2 \right]=\alpha_{mk}$, $\forall m$, exploiting the effect of channel hardening \cite{Ref_jiang2021cellfree}.
Thus, the downlink achievable rate for user $k$ is lower bounded by
\begin{equation}
    R_k^{CBF}= \log_2\left(1+\frac{ \left( \sum_{m=1}^M \sqrt{\eta_{mk}} \alpha_{mk} \right)^2}{\sigma_n^2/P_m+ \sum_{m=1}^M   \sum_{i=1}^K \eta_{mi} \beta_{mk}\alpha_{mi}}\right).
\end{equation}

Unlike maximizing the desired signals in CBF, ZFP completely suppresses multi-user interference (MUI) based on the downlink CSI. The first and second steps of the communication process in ZFP are the same as CBF, and then continues as
\begin{enumerate} \setcounter{enumi}{2}
\item Each AP reports its local CSI  so that the CPU gets the global CSI $\hat{\mathbf{G}}=[\hat{\mathbf{g}}_{1},\hat{\mathbf{g}}_{2},\ldots,\hat{\mathbf{g}}_{M} ]\in \mathbb{C}^{K\times M}$.
\item The CPU jointly precodes the information symbols in terms of \begin{equation}
    \textbf{s} = \hat{\mathbf{G}}^H\left(\hat{\mathbf{G}}\hat{\mathbf{G}}^H\right)^{-1} \boldsymbol{\eta} \textbf{u},
\end{equation} where $\boldsymbol{\eta} $ is a diagonal matrix consisting of power-control coefficients, i.e., $\boldsymbol{\eta}=\mathrm{diag}\{\sqrt{\eta_1},\ldots,\sqrt{\eta_K}\}$.
\item AP $m$ gets its transmitted symbol $s_m$ from the CPU, and then these APs synchronously transmit.
\end{enumerate}
Thus, the $k^{th}$ user observes
\begin{align} \nonumber \label{CFmMIMO:Eqn:theRXsignalofZF}
y_k &= \sqrt{P_m}\mathbf{g}_k \hat{\mathbf{G}}^H\left(\hat{\mathbf{G}}\hat{\mathbf{G}}^H\right)^{-1} \boldsymbol{\eta}\mathbf{u}+n_k\\ \nonumber
&= \sqrt{P_m}\left(\hat{\textbf{g}}_k+ \tilde{\textbf{g}}_k \right) \hat{\mathbf{G}}^H\left(\hat{\mathbf{G}}\hat{\mathbf{G}}^H\right)^{-1} \boldsymbol{\eta}\mathbf{u}+n_k\\
&= \underbrace{\sqrt{P_m\eta_k}u_k}_{\text{Desired Signal}} +  \underbrace{\sqrt{P_m} \tilde{\textbf{g}}_k  \hat{\mathbf{G}}^H\left(\hat{\mathbf{G}}\hat{\mathbf{G}}^H\right)^{-1} \boldsymbol{\eta} \mathbf{u}}_{\text{CSI Estimate Error}}+ n_k,
\end{align}
given the estimation error $\tilde{\mathbf{g}}_k=\mathbf{g}_k-\hat{\mathbf{g}}_k$, where $\mathbf{g}_k\in \mathbb{C}^{1\times M}$ is the actual CSI, which is the $k^{th}$ row of the channel matrix $\textbf{G}$, and $\hat{\mathbf{g}}_k\in \mathbb{C}^{1\times M}$ is an estimate of $\mathbf{g}_k$.
The achievable rate of ZFP user $k$ is lower bounded by \cite{Ref_nayebi2017precoding}
\begin{equation}
    R_k^{ZFP}= \log_2\left(1+\frac{\eta_k}{\sigma_n^2/P_m + \sum_{i=1}^K \eta_i \chi_k^i}\right),
\end{equation}
where $\chi_{k}^i$, $k=1,2,\ldots,K$ denotes the $i^{th}$ diagonal element of the $K\times K$ matrix for user $k$:
\begin{equation}
    \mathbb{E}\biggl[ \left(\hat{\mathbf{G}}\hat{\mathbf{G}}^H\right)^{-1} \hat{\mathbf{G}}\mathbb{E}\left[\tilde{\textbf{g}}_k^H\tilde{\textbf{g}}_k\right]  \hat{\mathbf{G}}^H\left(\hat{\mathbf{G}}\hat{\mathbf{G}}^H\right)^{-1}\biggr],
\end{equation}
where $\mathbb{E}\left[\tilde{\textbf{g}}_k^H\tilde{\textbf{g}}_k\right]$ is a diagonal matrix with $\epsilon_{mk}$ on its $m^{th}$ diagonal element, i.e.,
\begin{equation} \mathbb{E}\left[\tilde{\textbf{g}}_k^H\tilde{\textbf{g}}_k\right]=
    \begin{bmatrix}
        \epsilon_{1k} & 0 & \ldots& 0\\
        0& \epsilon_{2k} &  \ldots& 0\\
        \vdots & \vdots &\ddots &\vdots\\
        0&0&\ldots & \epsilon_{Mk}
    \end{bmatrix}.
\end{equation}

\section{Reconfigurable Intelligent Surface}

\begin{figure}[!t]
    \centering
    \includegraphics[width=0.38\textwidth]{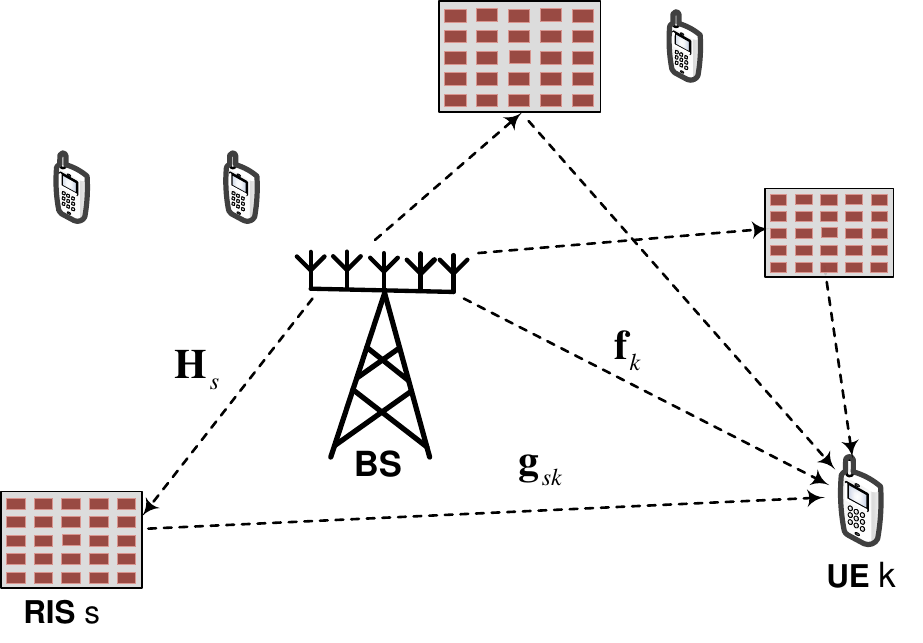}
    \caption{Illustration of a RIS-aided communications system.  }
    \label{fig:SystemModel}
\end{figure}

In a RIS-aided wireless system, an $M$-antenna BS serves $K$ users with the aid of $S\geqslant 1$ surfaces \cite{Ref_jiang2023performance}, as illustrated in \figurename \ref{fig:SystemModel}. Define the BS-UE, BS-RIS, and RIS-UE channels as $\mathbf{f}_{k}$, $\mathbf{H}_{s}$, and $\mathbf{g}_{sk}$, respectively, where each entry is in the form of $g=\sqrt{\beta}h$, as in CFmMIMO. The $s^{th}$ surface, $\forall s\in \{1,2,\ldots,S\}$, has $N_s$ elements, and $N=\sum_{s=1}^SN_s$ stands for the total number of elements. Write $\phi_{sn_s}\in [0,2\pi)$, $\forall s,n_s$ to denote the induced phase shift of a typical element $n_s\in \{1,2,\ldots,N_s\}$ over the $s^{th}$ surface. The key idea of RIS is to adjust $\phi_{sn_s}$, $\forall s, n_s$ adaptive to the instantaneous CSI such that all reflected signals are coherently combined at the receiver.
An element is not able to generate different phase shifts across  frequency (non-frequency-selective) but can be switched quickly in a short interval of \si{\micro\second}. As recommended by \cite{Ref_zheng2020intelligent_COML},  time-division multiple access (TDMA) is superior as multiple access for RIS. A radio frame is divided into $K$ time slots, and each user cyclically accesses its assigned slot.

At slot $k$, the BS applies linear beamforming $\mathbf{w}_k\in \mathbb{C}^{M\times 1}$ to send the information symbol $u_k$ intended for a general user $k$.
Define a \textit{time-selective} phase-shift matrix as
\begin{equation}
    \boldsymbol{\Phi}_s[k]=\mathrm{diag}\Bigl\{e^{j\phi_{s1}[k]},\ldots,e^{j\phi_{sN_s}[k]}\Bigr\},
\end{equation} where $\phi_{sn}[k]$ stands for a time-varying phase shift.
Thus, the received signal at the $k^{th}$ UE is given by
\begin{equation}
    y_k= \sqrt{P_d}\left(\sum_{s=1}^{S} \mathbf{g}_{sk}^T \boldsymbol{\Phi}_s[k] \mathbf{H}_s +\mathbf{f}_k^T\right)\mathbf{w}_k u_k +n_k,
\end{equation}
where $P_d$ expresses the power constraint of the BS.

Maximizing the received signal-to-noise ratio (SNR) of user $k$ needs the optimization \cite{Ref_jiang2022intelligent}
\begin{equation}
\begin{aligned} \label{eqnIRS:optimizationMRTvector}
\max_{\boldsymbol{\Phi}_s[k],\:\mathbf{w}_k}\quad &  \left|\left(\sum_{s=1}^S \mathbf{g}_{sk}^T \boldsymbol{\Phi}_s[k] \mathbf{H}_s +\mathbf{f}_k^T\right)\mathbf{w}_k\right|^2 \\
\textrm{s.t.} \quad & \|\mathbf{w}_k\|^2\leqslant 1\\
  \quad & \phi_{sn}[k]\in [0,2\pi), \: \forall s, n, k,
\end{aligned}
\end{equation}
which is non-convex since the objective function is not jointly concave with respect to $\boldsymbol{\Phi}_s[k]$ and $\mathbf{w}_k$. To solve this problem, we can apply alternating optimization \cite{Ref_wu2019intelligent}.
Given an initialized transmit vector $\mathbf{w}_k^{(0)}$, \eqref{eqnIRS:optimizationMRTvector} is simplified to
\begin{equation}  \label{eqnIRS:optimAO}
\begin{aligned} \max_{\boldsymbol{\Phi}_s[k]}\quad &  \left|\left(\sum_{s=1}^S\mathbf{g}_{sk}^T \boldsymbol{\Phi}_s[k] \mathbf{H}_s +\mathbf{f}_{k}^T\right)\mathbf{w}_k^{(0)}\right|^2\\
\textrm{s.t.}  \quad & \phi_{sn}[k]\in [0,2\pi), \: \forall s, n, k.
\end{aligned}
\end{equation}
The objective function enables a closed-form solution by applying the triangle inequality:
\begin{align} \nonumber
    &\left|\left(\sum_{s=1}^S\mathbf{g}_{sk}^T \boldsymbol{\Phi}_s[k] \mathbf{H}_s +\mathbf{f}_{k}^T\right)\mathbf{w}_k^{(0)}\right|\\ &\leqslant \left|\sum_{s=1}^S\mathbf{g}_{sk}^T \boldsymbol{\Phi}_s[k] \mathbf{H}_s \mathbf{w}_k^{(0)}\right| + \left|\mathbf{f}_{k}^T\mathbf{w}_k^{(0)}\right|,
\end{align}
achieving the equality if and only if
\begin{equation} \label{EQN_IRS_angleFormular}
    \arg\left (\sum_{s=1}^S\mathbf{g}_{sk}^T \boldsymbol{\Phi}_s[k] \mathbf{H}_s \mathbf{w}_k^{(0)}\right)= \arg\left(\mathbf{f}_{k}^T\mathbf{w}_k^{(0)}\right)\triangleq \varphi_{0k},
\end{equation}
where $\arg(\cdot)$ stands for the phase of a complex scalar.

Defining
\begin{align}
    &\mathbf{g}_k=\left[\mathbf{g}_{1k}^T,\mathbf{g}_{2k}^T,\ldots,\mathbf{g}_{Sk}^T \right]^T \\
    &\boldsymbol{\Phi}[k]=\mathrm{diag}\biggl\{\boldsymbol{\Phi}_1[k],\ldots,\boldsymbol{\Phi}_S[k]\biggr\}\\ \nonumber
    &\text{and}\\
    &\mathbf{H}=\Bigl[
\mathbf{H}_1^T, \mathbf{H}_2^T, \cdots, \mathbf{H}_S^T
\Bigr]^T,
\end{align}
 \eqref{EQN_IRS_angleFormular} is transformed to
\begin{equation}
    \arg\left (\mathbf{g}_{k}^T \boldsymbol{\Phi}[k] \mathbf{H} \mathbf{w}_k^{(0)}\right)= \arg\left(\mathbf{f}_{k}^T\mathbf{w}_k^{(0)}\right)\triangleq \varphi_{0k},
\end{equation}
Let $\mathbf{q}_k=\Bigl[q_{11}[k],\ldots,q_{SN_S}[k]\Bigr]^H$ with $q_{sn}[k]=e^{j\phi_{sn}[k]}$
and $\boldsymbol{\chi}_k=\mathrm{diag}(\mathbf{g}_k^T)\mathbf{H}\mathbf{w}_k^{(0)}$,
we have $\mathbf{g}_{k}^T \boldsymbol{\Phi}[k] \mathbf{H} \mathbf{w}_k^{(0)}= \mathbf{q}_k^H \boldsymbol{\chi}_k$.
Ignoring $\bigl|\mathbf{f}_{k}^T\mathbf{w}_k^{(0)}\bigr|$, \eqref{eqnIRS:optimAO} is transformed to
\begin{equation}  \label{eqnIRS:optimizationQ}
\begin{aligned} \max_{\boldsymbol{\mathbf{q}_k}}\quad &  \Bigl|\mathbf{q}_k^H\boldsymbol{\chi}_k\Bigl|\\
\textrm{s.t.}  \quad & |q_{sn}[k]|=1, \: \forall s,n,k,\\
  \quad & \arg(\mathbf{q}_k^H\boldsymbol{\chi}_k)=\varphi_{0k}.
\end{aligned}
\end{equation}
To achieve a coherent combining, solving \eqref{eqnIRS:optimizationQ} as
\begin{equation} \label{eqnIRScomplexityQ}
    \mathbf{q}^{(1)}_k=e^{j\left(\varphi_{0k}-\arg(\boldsymbol{\chi}_k)\right)}=e^{j\left(\varphi_{0k}-\arg\left( \mathrm{diag}(\mathbf{g}_k^T)\mathbf{H}\mathbf{w}_k^{(0)}\right)\right)}.
\end{equation}

Given $\boldsymbol{\Phi}^{(1)}[k]=\mathrm{diag}\left\{\mathbf{q}^{(1)}_k\right\}$, the BS can apply matched filtering to optimize the beamforming, resulting in $\mathbf{w}_k^{(1)} = \frac{\left(\mathbf{g}_{k}^T \boldsymbol{\Phi}^{(1)}[k] \mathbf{H} +\mathbf{f}_{k}^T\right)^H}{\left\|\mathbf{g}_{k}^T \boldsymbol{\Phi}^{(1)}[k] \mathbf{H} +\mathbf{f}_{k}^T\right\|}$.
The results of the first optimization iteration, i.e.,  $\boldsymbol{\Phi}^{(1)}[k]$ and $\mathbf{w}_k^{(1)}$, are inputted to the second iteration to derive $\boldsymbol{\Phi}^{(2)}[k]$ and $\mathbf{w}_k^{(2)}$.
This process iterates until the convergence is achieved with the optimal $\mathbf{w}_k^{\star}$ and  $\boldsymbol{\Phi}^{\star}[k]$, $\forall k$. Then, the achievable rate of user $k$ is computed by
\begin{align} \label{IRS_EQN_TDMA_SE} \nonumber
    R_k^{RIS}=\frac{1}{K}\log\left(1+\biggl \| \mathbf{g}_{k}^T \boldsymbol{\Phi}^\star[k] \mathbf{H} +\mathbf{f}_k^T \biggr\|^2 \frac{P_d}{\sigma_n^2} \right).
\end{align}

\section{Performance Comparison}
Performance comparisons between CFmMIMO and RIS in different scenarios and settings are conducted. We obtain the numerical results in terms of per-user spectral efficiency, as well as sum throughput, where the $5^{th}$ percentile SE, which is usually used to measure cell-edge performance, is highlighted.
\subsection{Simulation Setup}

Mimicking a realistic scenario, we consider a square area of $1\mathrm{km} \times 1\mathrm{km}$, over which CFmMIMO evenly distributes $M$ AP antennas and $K$ users. For RIS, a BS having M co-located antennas is installed at the center of this area while S RIS surfaces are randomly distributed. Large-scale fading is computed according to the formula $10^\frac{\mathcal{P}+\mathcal{X}}{10}$, where $\mathcal{P}$ denotes path loss and $\mathcal{X}$ is Log-Normal  shadowing fading, following $\mathcal{X}\sim \mathcal{N}(0,\sigma_{x}^2)$ with $\sigma_{x}=8\mathrm{dB}$. The COST-Hata model \cite{Ref_ngo2017cellfree} is applied to determine the path loss, i.e.,
\begin{equation} \label{eqn:CostHataModel}
    \mathcal{P}=
\begin{cases}
-\mathcal{P}_0-35\lg(d), &  d>d_1 \\
-\mathcal{P}_0-15\lg(d_1)-20\lg(d), &  d_0<d\leq d_1 \\
-\mathcal{P}_0-15\lg(d_1)-20\lg(d_0), &  d\leq d_0
\end{cases},
\end{equation}
where $d$ stands for the propagation distance.  The three-slope breakpoints take values $d_0=10\mathrm{m}$ and $d_1=50\mathrm{m}$, respectively, and the path loss at the 1-meter reference distance
\begin{IEEEeqnarray}{ll}
 \mathcal{P}_0=46.3&+33.9\log_{10}\left(f_c\right)-13.82\log_{10}\left(h_{tx}\right)\\ \nonumber
 &-\left[1.1\log_{10}(f_c)-0.7\right]h_{rx}+1.56\log_{10}\left(f_c\right)-0.8
\end{IEEEeqnarray} equals to $140.72\mathrm{dB}$ with carrier frequency $f_c=1.9\mathrm{GHz}$, the heights of transmitting and receiving antennas $h_{tx}=15\mathrm{m}$ and $h_{tx}=1.65\mathrm{m}$. Unlike moving users, a favorable location is deliberately selected for a RIS to capture the BS signal as well as possible. Usually, a line-of-sight (LOS) path between the BS and RIS exists. Consider a macro-cell scenario, where there is no reflector surrounding the BS and RIS, the signal transmission more likes free-space propagation, which can be computed by  $\frac{\mathcal{P}_0}{d^{-\alpha}}$,
where $\alpha$ means the path-loss exponent, set $\alpha=2.5$ in our simulations. Small-scale fading is assumed to be frequency flat and is modeled by a circularly-symmetric complex Gaussian random variable with zero mean and unit variance, i.e., $h \sim \mathcal{CN}(0, 1)$.

Conforming to the practical LTE and NR specifications, the BS power constraint is set to $P_d=20\mathrm{W}$. For a fair comparison under the same total power consumption, the transmit power per AP antenna in CFmMIMO $P_m=P_d/M$.  The power density of AWGN equals $-174\mathrm{dBm/Hz}$ with a noise figure of $9\mathrm{dB}$, and the signal bandwidth is $10\mathrm{MHz}$.  Power-control methods play an important role in CFmMIMO. Based on the criterion of using low-complexity schemes from a practical viewpoint, CBF and ZFP adopt the full-power transmission strategy  $\eta_{mk}=\left(\sum_{k=1}^{K_r} \alpha_{mk} \right)^{-1}$ and
$\eta_{1}=\ldots=\eta_{K}= \left( \max_m  \sum_{k=1}^{K} \delta_{km} \right )^{-1}$,
respectively, where
\begin{align}
\boldsymbol \delta_m & = \left[\delta_{1m},\ldots,\delta_{Km}\right]^T\\ \nonumber
&=\mathrm{diag}\left(\mathbb{E}\left[  \left(\hat{\mathbf{G}}\hat{\mathbf{G}}^H\right)^{-1}    \hat{\mathbf{g}}_m \hat{\mathbf{g}}_m^H   \left(\hat{\mathbf{G}}\hat{\mathbf{G}}^H\right)^{-1} \right]\right),
\end{align}
and $\hat{\mathbf{g}}_m$ denotes the $m^{th}$ column of $\hat{\mathbf{G}}$.

\subsection{Numerical Results}
\begin{figure}[!t]
\centering
\subfloat[]{
\includegraphics[width=0.4\textwidth]{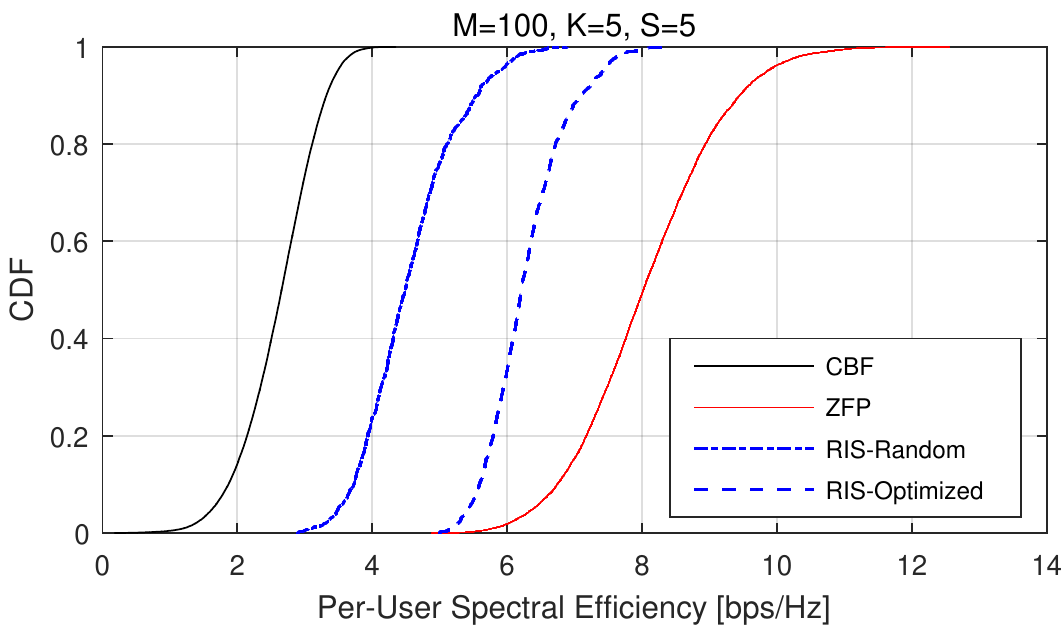}
\label{Fig_CDF_perUser}
}
\\
\subfloat[]{
\includegraphics[width=0.4\textwidth]{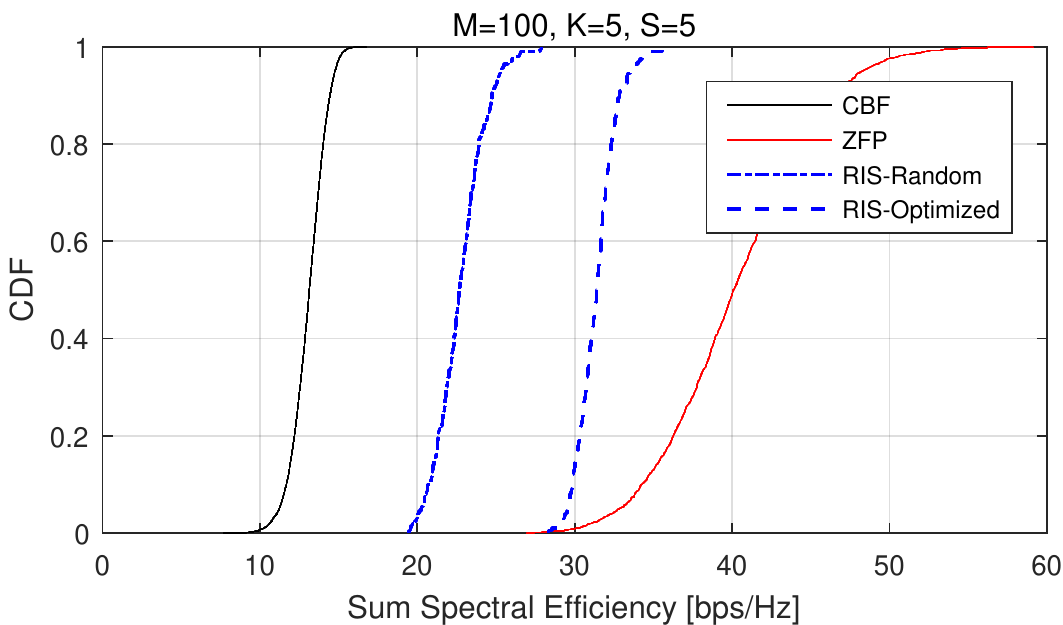}
\label{Fig_CDF_SC} \label{Fig_Sim01}
}
\caption{Performance comparison of CBF, ZFP, and RIS with optimized and random phase shifts in terms of (a) per-user SE and (b) sum throughput, where the BS equipping $M=100$ co-located or distributed antennas serves $K=5$ users, and $S=5$ RIS surfaces with $N_s=200$ reflecting elements per surface are applied in RIS-aided communications.  }
\end{figure}

We first obtain numerical results, as shown in Fig.3, in a typical scenario where a BS equipping $M=100$ co-located or distributed antennas serves $K=5$ users. In RIS-aided communications, $S=5$ surfaces with $N_s=200$ reflecting elements per surface are considered.  The acquisition of CSI in RIS is tough due to passive devices and a large number of reflected paths. To avoid the high complexity of RIS channel estimation, the phase shifts of RIS can be randomly set.  Hence, two modes are applied for RIS in our simulations: RIS with optimized phase shifts given in \eqref{eqnIRScomplexityQ} and RIS with random phase shifts. \figurename \eqref{Fig_CDF_perUser} gives the results in terms of cumulative distribution function (CDF) of per-user spectral efficiency. As expected, ZFP substantially outperforms CBF with the cost of high backhaul overhead and high computational complexity. ZFP achieves the $5^{th}$ percentile SE of \SI{6.4}{\bps\per\hertz^{}}, far greater than \SI{1.6}{\bps\per\hertz^{}} in CBF. RIS gets the $5^{th}$ percentile SE of \SI{3.5}{\bps\per\hertz^{}} and \SI{5.4}{\bps\per\hertz^{}} using random and optimized phase shifts, respectively. It enables flexibility for  a performance-complexity trade-off in a practical system design. In addition to per-user SE, the sum throughput for all five users is also obtained for reference, where CBF, ZFP, RIS with random and optimized phase shifts achieve \SI{13.1}{\bps\per\hertz^{}}, \SI{22.8}{\bps\per\hertz^{}}, \SI{31.4}{\bps\per\hertz^{}}, and \SI{40.3}{\bps\per\hertz^{}}, respectively.

\begin{figure}[!t]
    \centering
    \includegraphics[width=0.4\textwidth]{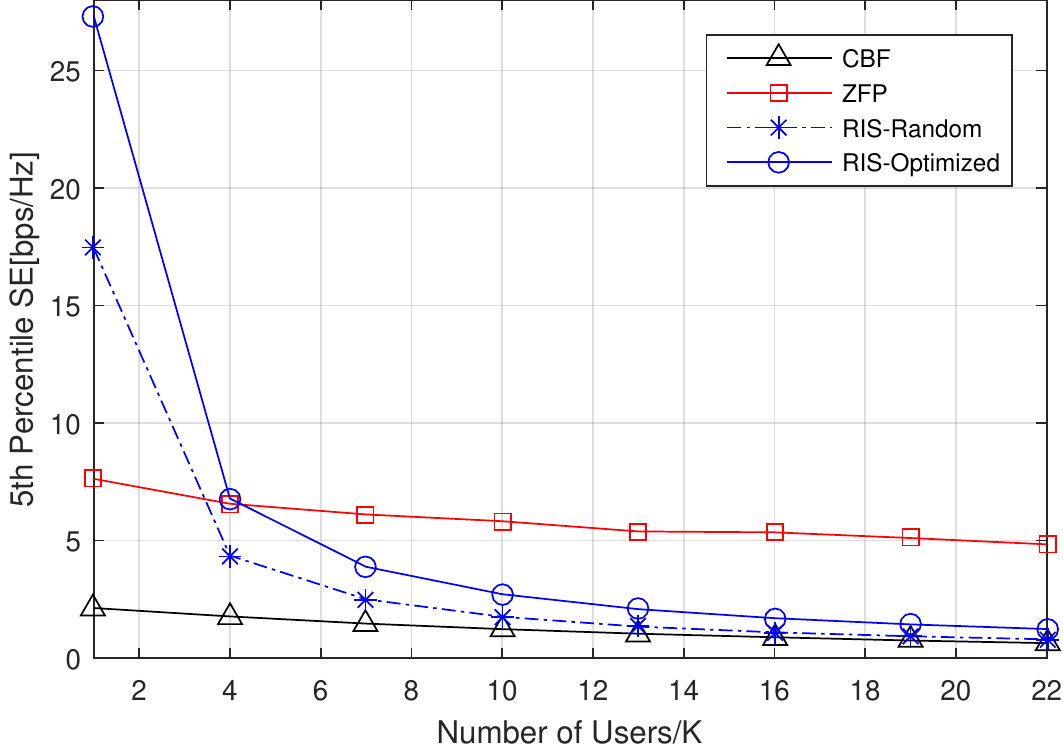}
    \caption{Performance comparison of 5th percentile spectral efficiency as a function of the number of users $K$. }
    \label{fig:PerUserK}
\end{figure}

Then, we further observe the behaviors of different techniques to deal with more users. \figurename \ref{fig:PerUserK} illustrates the $5^{th}$ percentile SE as a function of the number of users $K$ ranging from $1$ to $22$. As expected, the best result is achieved when there is only a single user in the system because all time-frequency resources are dedicated to this user and no MUI exists. The per-user performance decreases with the increasing number of users since different users have to share the resources. We observe that the per-user performance of CFmMIMO, either CBF or ZFP, is quite stable for user numbers. For example, the $5^{th}$ percentile SE of ZFP drops from around \SI{7.5}{\bps\per\hertz^{}} to \SI{5}{\bps\per\hertz^{}} when $K$ raises from $1$ to $22$. That is because the signals of all users are \textit{simultaneously} transmitted over the same time-frequency resource, and linear precoding at the BS can well handle inter-user interference. However, RIS is much more sensitive to the number of users since the  \textit{orthogonal} multiple access (i.e., TDMA) only allocates a small portion of the resource to each user, which is inversely proportional to the number of users, namely $1/K$. As shown in \figurename \ref{fig:PerUserK}, CFmMIMO with ZFP is superior to serving a larger number of users in terms of cell-edge performance.

\begin{figure}[!t]
\centering
\subfloat[]{
\includegraphics[width=0.4\textwidth]{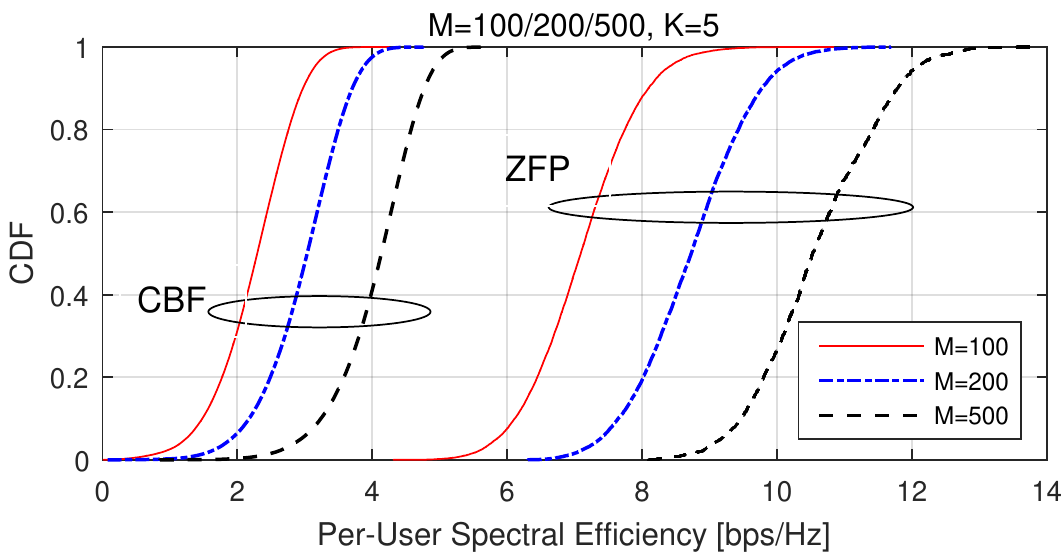}
\label{Fig_3a}
}
\\
\subfloat[]{
\includegraphics[width=0.4\textwidth]{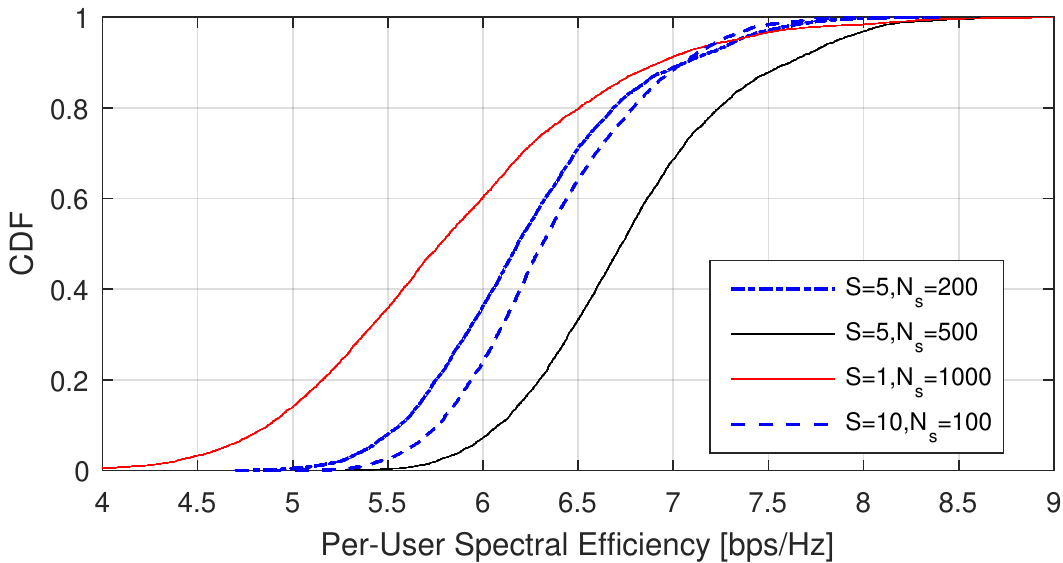}
\label{Fig_3b}
} \label{Fig_Sim03}
\caption{Performance comparison in terms of the CDF of per-user SE: (a) CFmMIMO with different numbers of APs $M=100$, $200$, and $500$;
 (b) RIS with optimized phase shifts under different numbers of surfaces $S=1, 5, 10$ and different numbers of elements per surface.  }
\end{figure}

In addition, we observe the impact of different numbers of APs in a CFmMIMO system, as shown in \figurename \ref{Fig_3a}. For a fair comparison, the total power consumption of CFmMIMO keeps the same, namely the transmit power per AP decreases accordingly. Increasing from $M=100$ to $M=500$ APs, per-user SE improves despite the same power consumption. That is because the density of APs becomes large, resulting in a smaller propagation distance between a typical user to APs, where the energy is not wasted to compensate for path loss. It implies that cell-edge performance can be improved by deploying more APs. But it turns to a higher deployment cost and high overhead of fronthaul. That is another performance-complexity trade-off to be considered in a practical system design. We also study the effects of
different numbers of surfaces and numbers of elements per surface on per-user SE using RIS. It is concluded that more reflecting elements and more distributed RIS bring a better result.

\section{Conclusions}
This paper compared two 6G-enabling technologies, i.e., reconfigurable intelligent surface and cell-free massive MIMO, from a particular perspective of cell-edge performance. We conducted simulations close to realistic scenarios and illustrated extensive numerical results to clarify their behaviors at the cell edge. Observing the $5^{th}$ percentile spectral efficiency, it can be concluded that both technologies can effectively boost cell-edge performance. The results provided some hints on the system design considering performance-complexity trade-offs, such as the antenna density in CFmMIMO and randomness in RIS phases.

\bibliographystyle{IEEEtran}
\bibliography{IEEEabrv,Ref_EuCNC2023}

\end{document}